\documentclass[11pt,a4paper]{article}
\usepackage{bm}
 \usepackage{a4wide}
\usepackage{latexsym}%
\usepackage{amsmath,amsfonts,amssymb}
\usepackage{graphicx,epsfig}
\usepackage{epstopdf}
\usepackage{amsthm}

\def\rmi{{\rm i}}
\def\rmd{{\rm d}}
\def\rme{{\rm e}}

\newsavebox{\uuunit}
\sbox{\uuunit}
    {\setlength{\unitlength}{0.825em}
     \begin{picture}(0.6,0.7)
        \thinlines
        \put(0,0){\line(1,0){0.5}}
        \put(0.15,0){\line(0,1){0.7}}
        \put(0.35,0){\line(0,1){0.8}}
       \multiput(0.3,0.8)(-0.04,-0.02){12}{\rule{0.5pt}{0.5pt}}
     \end {picture}}
\newcommand {\unity}{\mathord{\!\usebox{\uuunit}}}

\newcommand{\mathsym}[1]{{}}

\newcommand{\SO}{\mathop{\rm SO}}

\newcommand{\SL}{\mathop{\rm SL}}
\newcommand{\GL}{\mathop{\rm GL}}

\newcommand{\SU}{\mathop{\rm SU}}

\newcommand{\U}{\mathrm U}
\newcommand{\E}{\mathrm E}

\newtheorem{theorem}{Theorem}[section]

\begin{document}
\begin{flushright}
{\small
DFTT 37/2009\\
UUITP-10/09}

\normalsize
\end{flushright}

\begin{center}

{\LARGE \bf{The Full Integration of Black Hole Solutions
\\\vspace{0.3cm} to Symmetric
Supergravity Theories}}\\

\vspace{1cm} {\large  W.
Chemissany$^{\dag}$, J. Rosseel$^{\ddag}$,
M. Trigiante$^{\natural}$ and
T. Van Riet$^\flat$} \\[3mm]\vspace{2mm}
$\dag${\small\slshape University of Lethbridge, Physics Dept., \\
Lethbridge Alberta, Canada T1K 3M4
   }\\
{\upshape\ttfamily  wissam.chemissany@uleth.ca}\\[3mm]
$\ddag$
{\small\slshape Dipartimento di Fisica Teorica, Universit\`{a} di Torino \& INFN-Sezione di Torino,\\
Via P. Giuria 1, I-10125 Torino, Italy}\\
{\upshape\ttfamily  rosseel@to.infn.it }\\[3mm]
$\natural$
{\small\slshape Dipartimento di Fisica Politecnico di
Torino,\\
C.so Duca degli Abruzzi, 24, I-10129 Torino, Italy}\\
{\upshape\ttfamily   mario.trigiante@polito.it }\\[3mm]
$\flat$
{\small\slshape Institutionen f\"{o}r Fysik och Astronomi,\\
Box 803, SE-751 08 Uppsala, Sweden \\}
{\upshape\ttfamily  thomas.vanriet@fysast.uu.se}\\[15mm]

{\bf Abstract}
\end{center}

\begin{quotation}
\small

We prove that all stationary and spherical symmetric black hole
solutions to theories with symmetric target spaces are integrable
and we provide an explicit integration method. This exact
integration is based on the description of black hole solutions as
geodesic curves on the moduli space of the theory when reduced
over the time-like direction. These geodesic equations of motion
can be rewritten as a specific Lax pair equation for which
mathematicians have provided the integration algorithms when the
initial conditions are described by a diagonalizable Lax matrix.
On the other hand, solutions described by nilpotent Lax matrices, which originate from extremal regular (small)
$D=4$ black holes can be obtained as suitable limits of solutions obtained in the diagonalizable case,
 as we show on the
\emph{generating geodesic} (i.e. most general geodesic modulo
global symmetries of the $D=3$ model) corresponding to regular
(and small) $D=4$ black holes. As a byproduct of our analysis we
give the explicit form of the ``Wick rotation'' connecting the
orbits of BPS and non-BPS solutions in maximally supersymmetric
supergravity and its STU truncation.
\end{quotation}

\newpage

\pagestyle{plain} \tableofcontents


\section{Introduction}
The construction and study of black hole solutions in supergravity
has a long history.  Most of the research on this has focused on
extremal black holes, not necessarily  preserving supersymmetry
(see e.g. \cite{Mohaupt:2000gc, Andrianopoli:2006ub,
Pioline:2008zz, Ferrara:2008hwa} for reviews). The preservation of
supersymmetry makes an understanding of the string theory
microstates easier, while the vanishing of certain supersymmetry
variations implies that the second-order equations of motion can
be integrated to first-order equations, simplifying the explicit
construction of solutions. Recently it has been shown
\cite{Ceresole:2007wx, Andrianopoli:2007gt,
LopesCardoso:2007ky,Bellucci:2008sv,Gimon:2007mh} that similar
integrations to first-order equations, mimicking the supersymmetry
variations, can be carried out for some extremal
non-supersymmetric
\cite{Ortin:1997yn,Kallosh:2005ax,Bellucci:2006xz} and even some
non-extremal solutions \cite{Miller:2006ay, Perz:2008kh}.

In this note, we will follow a different approach to solve for
black hole solutions in supergravity, which does not use (hidden)
supersymmetry. We use the fact that the black hole solutions,
after performing a dimensional reduction of the supergravity
theory over the time direction, are described by geodesic curves
on a non-linear sigma model \cite{Breitenlohner:1987dg}, see also
\cite{Gal'tsov:1998yu,Clement:1986bt,Clement:1996nh,Clement:1985gm}
for early original work. This idea has recently been fleshed out
in more detail to understand the general structure of BPS
solutions \cite{Bossard:2009at, Gunaydin:2005mx}, non-BPS
attractors \cite{Gaiotto:2007ag}, or the general properties of
extremal and non-extremal solutions \cite{Bergshoeff:2008be}. In
case the sigma model is described by a symmetric space $G/H$, the
geodesic curves are classified in terms of the Noether charges of
the solution. In particular, when the coset representative
$\mathbb{L}$ is squared to the symmetric coset matrix,
$\mathcal{M}=\mathbb{L}\mathbb{L}^T$ \footnote{$\mathbb{L}^T$
denotes the generalised transpose and not the ordinary transpose,
see for instance \cite{Bergshoeff:2008be}.}, then the solution can
be compactly written as \cite{Breitenlohner:1987dg}
\begin{equation}
\mathcal{M}(t)=\mathcal{M}(0)\,\rme^{Q\,t}\,,
\end{equation}
where $Q$ is a matrix containing the Noether charges. This can be
seen as a proof of principle that the geodesic equations are
integrable\footnote{Proving Liouville integrability of the
Hamiltonian system associated with the $D=3$ model is a subtler
issue. It requires the knowledge of a number of conserved
quantities in involution equal to the number of scalar fields.
This problem will be dealt with elsewhere.}. From the knowledge of
$Q$ one can extract already a lot of useful information about the
black hole solutions \cite{Bossard:2009at,Gunaydin:2005mx,
Gaiotto:2007ag, Bergshoeff:2008be}, but in many cases more is
needed to understand the physics of the black hole solutions. For
instance, the scalar $U$ describing the black hole geometry is
contained in $\mathcal{M}$ and an explicit understanding of the
black hole geometry requires the full radial dependence of this
scalar. However, extracting the expressions for the scalars out of
the expression for $\mathcal{M}$ is a hard problem; hence there is
the need for an explicit integration method on the level of the
scalars.\par The integration procedure we will use has first been
introduced in supergravity in constructing time-dependent
solutions \cite{Fre:2003ep,  Fre:2005bs, Fre':2005sr, Fre':2007hd,
Fre:2008zd}, for which integration through supersymmetry is absent
from the very beginning. If the solutions possess Killing
directions, they are described by geodesics on a sigma model of a
lower-dimensional supergravity, similar to black
holes\footnote{See \cite{Bergshoeff:2008be} for a general unifying
explanation of this principle.}. The difference between
time-dependent solutions and stationary solutions is that in the
former case the geodesics live in a Riemannian coset, while in the
latter case they live in a coset of indefinite signature. This
difference matters when constructing solutions and a main
characteristic is that the pseudo-Riemannian case is richer and
more involved. In the Riemannian case, it was first pointed out in
\cite{Fre:2005bs} that the geodesic equations could be rewritten
as a Lax pair and that an explicit integration procedure had been
developed by mathematicians \cite{kodama2-1995}. This was recently
used to construct very non-trivial time-dependent solutions in
supergravity \cite{Fre:2008zd}.\par One of  the main purposes of
this paper is to demonstrate that this holds also in the
pseudo-Riemannian case. Namely, the geodesic equations can again
be written as a Lax pair and the Lax pair is again of a very
specific form for which an explicit integration algorithm has been
worked out in the mathematical literature \cite{kodama3-1995}
\emph{assuming that the initial conditions, summarized by giving
the Lax matrix at some initial time, are diagonalizable}. The
latter restriction might seem rather unimportant since
non-diagonalizable matrices are a subset of measure zero in the
space of matrices. However, ironically,  all BPS states (and
non-BPS attractive black holes) are described by such initial
conditions.  It is rather curious that exactly non-extremal
solutions are easier to describe than extremal solutions in this
Lax pair approach. This demonstrates that the geodesic approach is
orthogonal to using (fake) supersymmetry \footnote{We refer to
\cite{Perz:2008kh} for a comparison of the two approaches.}. In
this paper we will not yet develop the general Lax integration
algorithm for non-diagonalizable initial conditions; we will
however already mention that the Lax integration algorithm can be
extended to the non-diagonal case. We leave a discussion on this
issue for an upcoming paper that will contain more technical
details \cite{Chemissany}. Here, we simply want to put forward the
principle, and illustrate this with simple examples based on the
cosets $\SL(2,\mathbb{R})/\SO(1,1)$ and
$\SL(3,\mathbb{R})/\SO(2,1)$. A second part of our analysis
concerns the application to the description of  $D=4$ black holes.
As we shall show, a class of solutions with non-diagonalizable
initial conditions, which can be obtained as limits of solutions
with diagonalizable Lax matrices, and which therefore can be
derived  by our algorithm, are precisely those which are relevant
to the description of non-singular (i.e. having no naked
singularities) $D=4$ black hole solutions. We shall apply our
algorithm to the construction of the \emph{generating geodesic} of
regular (and small) $D=4$ black holes. By generating geodesic we
mean the solution to the $D=3$ model which depends on the least
number of parameters such that, by applying on it the global
symmetries of the model, the most general geodesic can be
constructed. Such geodesic unfolds in a simpler submanifold of the
scalar manifold containing a characteristic number of $dS_2$
factors times ${\rm O}(1,1)$ factors. This solution was originally
studied in \cite{Bergshoeff:2008be}. Here, we shall further
elaborate on it and, as a byproduct, we determine the explicit
form of the ``Wick rotation'' connecting the orbits of BPS and
non-BPS solutions in maximally supersymmetric supergravity and its
STU truncation.
\par The outline of the paper is as follows. In section
\ref{sec:eomisLax} we show that the geodesic equations, that
describe black hole solutions in supergravities with symmetric
target spaces, can be rewritten in Lax pair form. In section
\ref{sec:laxalgsgen} we give a summary of the integration
algorithm that allows us to integrate the geodesic equations on
pseudo-Riemannian symmetric target spaces in Lax pair form. This
algorithm will be illustrated in section \ref{sec:ex}, using the
examples of $\SL(2,\mathbb{R})/\SO(1,1)$ and
$\SL(3,\mathbb{R})/\SO(2,1)$. In section \ref{ndic}, after
recalling some results of \cite{Bergshoeff:2008be}, we shall make
general comments on the solutions corresponding to (non-)diagonalizable initial conditions and show that the generating
geodesic with non-diagonalizable initial conditions can be
obtained as a limit of solutions with diagonalizable Lax matrices.
As an example of solutions with non-diagonalizable initial
conditions, the generating geodesic corresponding to extremal
$D=4$ regular (and small) black holes will be explicitly
constructed in the maximally supersymmetric theory and its STU
truncation. Our analysis will show that the Noether charge
matrices corresponding to such solutions belong, according to
their supersymmetry properties,  to different real sections of the
same orbit of the complexification of the global symmetry group.
Finally, in section \ref{sec:concl} we present our conclusions.
\par
In the final stage of preparation of the present paper, we have
learned about the interesting paper \cite{Fre:2009et} whose
results partially overlap ours.

\section{The geodesic equations in Lax pair form} \label{sec:eomisLax}

As mentioned in the introduction, in describing cosmological and
black hole solutions in supergravity, one often uses the existence
of certain Killing vectors in order to reduce the supergravity to
a lower dimension. The solutions are then essentially described by
geodesics on the non-linear sigma model, spanned by the scalar
fields of the lower-dimensional theory
\cite{Breitenlohner:1987dg}. This non-linear sigma model can be
either pseudo-Riemannian or Riemannian, depending on whether the
reduction to three dimensions includes the time direction or not.
In the following, we will consider the case in which the
non-linear sigma model is a symmetric space $G/H$, that can be
either Riemannian or pseudo-Riemannian. In the former case, $H$ is
the maximally compact subgroup of $G$, while in the latter case
$H$ is a non-compact subgroup of $G$. In this section, we will
show that the geodesic equations for the scalar fields can be
rewritten in Lax pair form, establishing their integrability,
irrespective of whether $G/H$ is Riemannian or pseudo-Riemannian.
The argument proceeds along the same lines as in
\cite{Fre:2005bs}, where $G/H$ was supposed to be Riemannian.

Let $\mathbb{G}$, $\mathbb{H}$  be the Lie algebras of $G$ and of
the (not necessarily compact) isotropy group $H$  respectively and
let $\mathbb{K}$ be the  orthogonal complement of $\mathbb{H}$ in
$\mathbb{G}$, as determined by the Cartan decomposition:
\begin{equation}
\mathbb{G} = \mathbb{H} + \mathbb{K} \,.
\end{equation}
This decomposition is defined through the use of the Cartan
involutive automorphism $\theta$, which acts as
\begin{equation}
\theta(\mathbb{H})=\mathbb{H}\,,\qquad
\theta(\mathbb{K})=-\mathbb{K} \,.
\end{equation}
Since the automorphism preserves the Lie bracket, we have
\begin{equation}\label{commutation}
[\mathbb{H},\mathbb{H}]\subset
\mathbb{H}\,,\qquad[\mathbb{H},\mathbb{K}]\subset\mathbb{K}\,,\qquad
[\mathbb{K}, \mathbb{K}]\subset \mathbb{H}\,.
\end{equation}
Let us illustrate this in case $G/H$ is a maximally non-compact
coset, i.e. $\mathbb{G}$ is the split real form of a complex Lie
algebra $\mathbb{G}_{\mathbb{C}}$. In that case, we have
\begin{align} \label{HKsplit}
\mathbb{H}&=\text{Span}\{t_{\alpha}\}\equiv\text{Span}\{
E^{\alpha}+\theta(E^{\alpha}) \} \nonumber \,,\\
\mathbb{K}&=\text{Span}\{K_A\}\equiv\text{Span}\{H_i,
\tfrac{1}{\sqrt{2}}\left(E^{\alpha}-\theta(E^{\alpha})\right)
\}\,,
\end{align}
where $H_i$ and $E^{\alpha}$ are the Cartan  and the positive step
operators (i.e. corresponding to positive roots) of the algebra
$\mathbb{G}$ respectively. For the cosets that originate from a
purely space-like reduction or from a reduction along the time
direction, the action of $\theta$ on the step operators is
respectively given by \cite{Bergshoeff:2008be}:
\begin{align} \label{acttheta}
 \text{purely space-like reduction}&:\theta(E^{\alpha})=-E^{-\alpha} \nonumber \,,\\
 \text{reduction including time }&:\theta(E^{\alpha})= -(-1)^{\beta_0(\alpha)}E^{-\alpha}
\end{align}
where $\beta_0(\alpha)=H_{\beta_0}(E^{\alpha})$ represents the
grading of the root $\alpha$ with respect to $H_{\beta_0}$, the
Cartan generator that is associated with the internal time
direction\footnote{This generator is normalized so that its
adjoint action on $\mathbb{G}$ has eigenvalues: 0 (on the
generators of the four-dimensional isometries and on $H_{\beta_0}$
itself), $\pm 1$ ($+1$ on the shift generators parametrized by the
internal component of the four-dimensional vector fields and the
scalars dual to the $D=3$ vectors; $-1$ on the shift generators
associated with the corresponding  negative roots); $\pm 2$ (on
the  generators $E_{\pm \beta_0}$ which, together with
$H_{\beta_0}$ generate the Ehlers $\SL(2,\mathbb{R})$,
$E_{\beta_0}$ being parametrized by the axion dual to the Kaluza
Klein vector of the  $D=4\rightarrow D=3$ reduction).}.
Let us now denote by $\mathbb{L}$ a coset representative. We wish
to write the geodesic flow equations in a Lax pair form both in
the Riemannian and pseudo-Riemannian cases. We shall generically
denote by $t$ the affine parameter along the geodesic, so that the
solution will be described by a suitable dependence of the $D=3$
scalar fields on $t$: $\phi^I=\phi^I(t)$ \footnote{When studying
$D=4$ spherically symmetric black holes, $t$ will be related to
the radial coordinate in the Euclidean $D=3$ theory originating
from a time-like reduction of the four-dimensional one; when
studying cosmological solutions $t$ will denote the time
coordinate in the Lorentzian $D=3$ theory arising from $D=4$
through a space-like reduction.}. The left invariant one-form
$\Omega$ on the coset, pulled-back on the geodesic, can be
expanded as follows
\begin{equation} \label{maurercartan}
\Omega=\mathbb{L}^{-1}\frac{\rmd}{\rmd
t}\mathbb{L}=\dot{\phi}^I\,\mathbb{L}^{-1}\frac{\partial}{\partial
\phi^I}\mathbb{L}= W^{\alpha}t_{\alpha} + V^A K_A \equiv W+V\,.
\end{equation}
We identify $V$ as the coset vielbein pulled-back to the
one-dimensional space parametrized by the affine parameter $t$.
The geodesic action reads
\begin{equation}\label{action}
S=\int \rmd t \,\text{Tr}(VV) \propto \int \rmd t
\,G_{IJ}(\phi)\,\dot{\phi}^I\dot{\phi}^J\,,
\end{equation}
where the trace is defined in some linear representation of $G$
and where we introduced coordinates (scalar fields) on the coset
via $V=K_A V^A_I\dot{\phi}^I$. To compute the equations of motion
we consider a variation of the action (\ref{action})
\begin{equation}
\delta S=2\int\rmd t\,\text{Tr}[ V \delta V]\,.
\end{equation}
The variation $\delta V$ can be rewritten by using the following
identity
\begin{equation}\label{deltaOmega}
\delta \Omega\equiv \delta W+\delta
V=[\Omega,\mathbb{L}^{-1}\delta\mathbb{L}] + \frac{\rmd}{\rmd
t}(\mathbb{L}^{-1}\delta\mathbb{L})\,.
\end{equation}
The functional $\mathbb{L}^{-1}\delta\mathbb{L}$ can then formally
be rewritten using the Cartan decomposition
\begin{equation}
\mathbb{L}^{-1}\delta \mathbb{L}=\delta w+\delta v
\end{equation}
where $\delta w$ and $\delta v$ represent the projection of
$\mathbb{L}^{-1}\delta \mathbb{L}$  on $\mathbb{H}$, $\mathbb{K}$
respectively. Plugging this in equation (\ref{deltaOmega}), and
projecting the resulting equation onto the $\mathbb{K}$ subspace
(using the commutation relations (\ref{commutation})) we find
\begin{equation}
\delta V=[W,\delta v] + [V,\delta w] + \frac{\rmd}{\rmd t}\delta
v\,.
\end{equation}
Finally, upon substituting this expression in the variation of the
action and using the cyclicity of the trace, we find
\begin{equation}
\delta S=2\int\rmd t\,\text{Tr}\Bigl[\bigl([V,W]-\frac{\rmd}{\rmd
t}V\bigr)\delta v\Bigr] \,,
\end{equation}
from which the Lax pair equation follows
\begin{equation}
\frac{\rmd}{\rmd t}V=[V,W]\,.\label{lax0}
\end{equation}
In the following, we will always assume that we work with a coset
representative $\mathbb{L}$ in solvable gauge. As explained in
e.g. \cite{Fre:2003ep,Fre:2005bs,Fre':2005sr,Fre':2007hd}, this
gauge is such that
\begin{equation}
W = V_{>0} - V_{<0} \,,
\end{equation}
where $V_{>0(<0)}$ denotes the upper-triangular (resp.
lower-triangular) part of $V$ \footnote{It is a consequence of Lie's
theorem that the generators $T_I$ of the solvable Lie group
describing a local patch on $G/H$ can be all represented, in a
suitable basis, by upper (or lower) triangular matrices. For
maximally non-compact cosets $G/H$, the solvability condition can be
written as $V^\alpha = \sqrt{2} W^\alpha$. From (\ref{HKsplit}),
(\ref{acttheta}) and (\ref{maurercartan}), one can then infer that
$W=V_{>0} - V_{<0}$.}. Note that for pseudo-Riemannian cosets, there
is a subtlety in choosing the solvable gauge. This gauge can not in
general be chosen globally on the manifold for non-compact $H$. The
solvable group defined by the  Iwasawa decomposition of $G$ with
respect to its maximal compact subgroup describes local patches of
the manifold. Only one of these patches is to be considered as
\emph{physical}, namely spanned by the physical fields of the
theory. At the boundary of this region some of these fields explode,
signalling singularities in the corresponding four dimensional
solution. In the physical solvable patch, time-like and null
geodesics (originating from regular four dimensional solutions) are
complete, while space-like are not, as they reach the boundary at a
finite value of the "proper time''. We will explicitly show this in
section \ref{ssec:sl2} where we review the simple case in which the
 scalar manifold $G/H$ is the  two-dimensional de Sitter
($dS_2$) space-time. This example will be particularly instructive
since, as will be shown in Section \ref{ndic}, the generating
geodesic (with respect to the action of $G$) of regular black
holes in $D=4$ is described as a geodesic in a product of $dS_2$
spaces.\par
%
Before starting the discussion of the integration algorithm for
the Lax pair equation, let us give the expression of the Noether
charge matrix in terms of the Lax operator and the coset
representative. Using (\ref{maurercartan}) and (\ref{lax0}) it is
straightforward to show that the following matrix
\begin{eqnarray}
Q&=&2\, \mathbb{L}^{-1\, T}\,V^T\,\mathbb{L}^{T}\,,\label{QV}
\end{eqnarray}
is a constant of motion. It encodes the conserved charges associated
with the invariance with respect to the left action of $G$. Let us
stress here that $Q$ is an object of $\mathbb{G}$ while $V$, being
in $\mathbb{K}$ only transforms under $H$. The action of a global
$G$-transformation on $Q$ decomposes into the action of a global
$G/H$ transformation on $\mathbb{L}$ whose effect is to move the
initial point of the geodesic, and the action of a global
$H$-transformation on $V$.
\section{The Lax algorithm} \label{sec:laxalgsgen}

In this section, we will consider an algorithm that is useful in
solving differential equations that can be written in Lax pair
form:
\begin{equation} \label{laxeq}
\frac{\rmd V}{\rmd t} = [V,W] \,,
\end{equation}
where $V$ and $W$ are $N\times N$-matrices and $W$ is given in
terms of $V$ as
\begin{equation} \label{defW}
W = V_{>0} - V_{<0} \,.
\end{equation}

For the problem of solving the geodesic equations on symmetric
spaces, the Lax operator $V$ is more specifically defined by
\begin{equation}
V(t) = \sum_A \mathrm{Tr}\left(\mathbb{L}^{-1} \frac{\rmd
\mathbb{L}}{\rmd t} K_A\right) K_A \,,
\end{equation}
where $K_A$ denote the generators of $\mathbb{K}$. As was shown in
section \ref{sec:eomisLax}, with this definition the Lax equation
(\ref{laxeq}) reproduces the geodesic equations on the symmetric
space $G/H$, irrespective of whether $G/H$ is Riemannian or
pseudo-Riemannian.

Depending on the symmetry properties of the Lax operator $V$,
algorithms have been devised that solve the matrix differential
equation (\ref{laxeq}) and lead to an explicit $t$-dependent
solution for $V$. After an explicit solution $V_{\mathrm{sol}}(t)$
for the Lax operator $V$ has been found, one can generically solve
for the scalars that parametrize the coset manifold, by solving
the following system of differential equations:
\begin{equation}
\mathrm{Tr}\left(\mathbb{L}^{-1} \frac{\rmd \mathbb{L}}{\rmd t}
K_A\right) = \mathrm{Tr} \left(V_{\mathrm{sol}}(t) K_A\right) \,.
\end{equation}
As the left-hand-side of these equations depends on the first
derivatives of the scalars, this is a system of first-order
equations. Depending on the specific parametrization used for the
coset representative, one can solve this system in an iterative
manner. The main difference between the Riemannian and the
pseudo-Riemannian case lies in the symmetry properties of the Lax
operator $V$. For Riemannian cosets, one can choose a matrix
representation of the Lie algebra, in which all generators $K_A$
are symmetric matrices. The Lax operator $V$ is then also given by
a symmetric matrix. The algorithm that solves equations
(\ref{laxeq}) was constructed in \cite{kodama2-1995,kodama-1995}.
For pseudo-Riemannian cosets, the generators $K_A$ are in general
no longer symmetric. Instead, some of these generators will be
symmetric (corresponding to the positive signature directions of
the coset), while others will be anti-symmetric matrices
(corresponding to the negative signature directions). Also the Lax
operator will therefore no longer be a symmetric matrix and will
in general be neither symmetric nor anti-symmetric. As was shown
in \cite{kodama3-1995}, the algorithm of
\cite{kodama2-1995,kodama-1995} can be extended to include also
this case. In the following, we will summarize this extended
algorithm. The algorithm for Riemannian spaces can then be found
as a special case of the one outlined below.

In \cite{kodama3-1995} an integration algorithm for the Lax
equations (\ref{laxeq}) (with $W$ given by (\ref{defW})) is
outlined, for Lax operators $V$ for which there exists a
non-degenerate diagonal matrix $S = \mathrm{diag}(s_1 \cdots
s_N)$, such that
\begin{equation}
\tilde{V} = V S \qquad \mathrm{is}\ \mathrm{a}\
\mathrm{symmetric}\ \mathrm{matrix}\,.
\end{equation}
The fact that this integration algorithm allows us to integrate
the geodesic equations for pseudo-Riemannian cosets is then a
result of the following theorem:
\begin{theorem} \label{theoremlor}
For pseudo-Riemannian symmetric spaces, one can always find a
suitable space of coset generators $\mathbb{K}$, such that there
exists a non-degenerate diagonal matrix $S$ with the property that
\begin{equation}
\forall K_A  \in \mathbb{K} \quad : \quad K_A S = (K_A S)^T  \,.
\end{equation}
\end{theorem}
In order to justify this statement, we note that one can always
find a linear representation for which the Cartan involution can
be written as
\begin{equation}
\theta(X) = - \eta X^T \eta \,,
\end{equation}
where $\eta$ is some diagonal matrix that squares to unity. In
general, this matrix $\eta$ is given by
\begin{equation}
\eta = \mathrm{diag}(-\unity_p,\unity_q)\,,
\end{equation}
for some $p$, $q$. As $\theta(\mathbb{K}) = -\mathbb{K}$, we thus
find that
\begin{equation}
K_A \eta = (K_A \eta)^T\,,
\end{equation}
for $K_A \in \mathbb{K}$, implying that one can take $S = \eta$.
Note that also Riemannian spaces obey this theorem. Indeed, in
that case $S=\unity$. The formulas that will be given below can
then be easily adapted to the Riemannian case by taking $s_i = 1$,
$\forall i$.

The algorithm itself is an instance of the inverse scattering
method and as such constructs the solution $V_{\mathrm{sol}}(t)$
for the Lax operator, starting from the initial conditions
contained in the Lax operator $V$ at $t=0$. The first step in
establishing the Lax algorithm consists in realizing that the Lax
operator can be diagonalized:
\begin{eqnarray}
V \Phi & = & \Phi \Lambda \,, \nonumber \\
\frac{\rmd}{\rmd t} \Phi & = & W \Phi \,,
\end{eqnarray}
where $\Lambda = \mathrm{diag}(\lambda_1, \cdots, \lambda_N)$ is
the diagonal matrix containing the eigenvalues of $V$. The matrix
$\Phi$ then contains the eigenvectors of $V$ as its columns. We
will denote the eigenvector of $V$ with eigenvalue $\lambda_k$ by
\begin{equation}
\phi(\lambda_k) = \left(\begin{array}{c} \phi_1(\lambda_k) \\
\vdots \\ \phi_N(\lambda_k) \end{array} \right) \,.
\end{equation}
The matrix $\Phi$ is thus given by
\begin{equation}
\Phi = \left[ \phi(\lambda_1) \cdots \phi(\lambda_N) \right] =
\left[\phi_i(\lambda_j)\right]_{1 \leq i,j \leq N} \,.
\end{equation}
One can moreover show that $\Phi$ can be chosen such that it obeys
the following orthogonality relations:
\begin{equation}
\Phi S^{-1} \Phi^T = S^{-1} \,, \qquad \Phi^T S \Phi = S \,.
\end{equation}
Note that in general $\Lambda$ can be complex. Also $\Phi$ will in
general be complex, even if $\Lambda$ is real. Only in the
Riemannian case will $\Lambda$ and $\Phi$ be real.

The $t$-dependent solutions for the matrix elements of $\Phi$ are
then given by
\begin{equation} \label{solphi}
\phi_i(\lambda_k, t) = \frac{\rme^{-\lambda_k t}}{\sqrt{D_i(t)
D_{i-1}(t)}} \left| \begin{array}{ccc} s_1 c_{11} & \cdots & s_1
c_{1i} \\ \vdots & \ddots & \vdots \\ s_{i-1} c_{i-1,1} & \cdots &
s_{i-1} c_{i-1,i} \\ \phi^0_1(\lambda_k) & \cdots &
\phi^0_i(\lambda_k) \end{array} \right| \,.
\end{equation}
The quantities $\phi^0$ represent the matrix $\Phi$ at $t=0$; i.e.
they are obtained from the eigenvalue problem
\begin{eqnarray}
& & V_0 \Phi^0 = \Phi^0 \Lambda \,, \nonumber \\
& & \Phi^0 = \left[ \phi^0_i(\lambda_j) \right]_{1 \leq i,j \leq
N} \,,
\end{eqnarray}
where $V_0 = V(0)$. Note that the eigenvalues contained in
$\Lambda$ are time-independent, a property often denoted as the
iso-spectral property of the Lax operator. The quantities
$c_{ij}(t)$ in the formula (\ref{solphi}) are given by
\begin{equation}
c_{ij}(t) = \sum_{k=1}^N s_{k}^{-1}\, \rme^{-2 \lambda_k t}\,
\phi^0_i(\lambda_k) \phi^0_j(\lambda_k) \,.
\end{equation}
The $D_k(t)$ are then given by the determinant of the $k\times
k$-matrix with entries $s_i c_{ij}(t)$, i.e.:
\begin{equation}
D_k(t) = \mathrm{det}\left[ \left( s_i c_{ij}(t) \right)_{1\leq
i,j \leq k} \right] \,.
\end{equation}
Note that $s_i c_{ij}(0) = \delta_{ij}$ and $D_k(0) = 1$.
Furthermore $D_0(t) = 1$.

The final solution for the Lax operator is then found as:
\begin{equation} \label{solLax}
\left[V_{\mathrm{sol}}(t)\right]_{ij} = s_j \sum_{k=1}^N s_k^{-1}
\lambda_k \phi_i(\lambda_k,t) \phi_j(\lambda_k,t)  \,.
\end{equation}

\section{Examples} \label{sec:ex}
In this section, we will illustrate the previously outlined
algorithm using two examples. The choice of the examples is both
based on simplicity as well as on physical relevance. The first
example deals with the $\SL(2,\mathbb{R})/\SO(1,1)$ coset, where
the geodesics can be found in a closed and rather simple form. We
will see that the algorithm indeed leads to the expected results.
The second example deals with the $\SL(3,\mathbb{R})/\SO(2,1)$
coset. Although still simple, this example is also physically
relevant, as it can for instance be used to find black hole
solutions in four-dimensional Einstein-Maxwell-dilaton theories.
This can be done by making use of the \emph{$4D$ black holes/ $3D$
$\SL(3,\mathbb{R})/\SO(2,1)-$geodesics correspondence }outlined in
the introduction. In this paper, we will restrict ourselves to
showing how geodesics on $\SL(3,\mathbb{R})/\SO(2,1)$ can be
produced using the Lax algorithm. The connection between these
geodesics and four- dimensional black holes will be worked out
more explicitly in a forthcoming paper.

\subsection{The $\SL(2,\mathbb{R})/\SO(1,1)$ example}
\label{ssec:sl2}

\begin{figure}[ht]
\begin{center}
\centerline{\includegraphics[width=8cm,height=7cm]{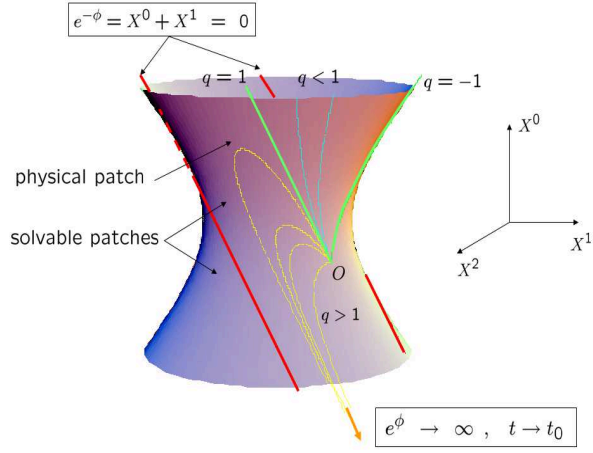}}
\caption{The $dS_2$ space-time in $\mathbb{R}^{1,2}$. Some geodesics
are represented, for $t<0$. The behavior for $t>0$ is obtained by
reflection with respect to the plane containing the $X^0$ axis and
passing through $O$. Time-like ($|q|<1$) and light-like ($|q|=1$)
geodesics are complete, space-like ones are not ($|q|>1$).
}\label{fig1}
\end{center}
\end{figure}

Let us apply the algorithm to the pseudo-Riemannian coset
$\SL(2,\mathbb{R})/\SO(1,1)$. The Cartan generator $H$ and positive
root $E$ are taken to be
\begin{equation}
H = \left(\begin{array}{cc} 1 & 0 \\ 0 & -1 \end{array}
\right)=\eta\,\,,\,\,\,E = \left(\begin{array}{cc} 0 & 1 \\ 0 & 0
\end{array} \right) \,.
\end{equation}
The Cartan decomposition $\mathbb{G} = \mathbb{H} + \mathbb{K}$ is
then determined by
\begin{equation}
\mathbb{H} = \mathrm{Span} \left\{\frac{1}{2}\,(E + E^T) \right\}
\,, \mathbb{K} = \mathrm{Span} \left\{H, \frac{1}{2}\,(E -
E^T)\right\} \,.
\end{equation}
 This is the two-dimensional de Sitter
space-time ($dS_2$) which can be represented by the following
hyperboloid in $\mathbb{R}^{1,2}$, see Figure \ref{fig1}:
\begin{eqnarray}
-(X^0)^2+(X^1)^2+(X^2)^2&=&\alpha^2\,,
\end{eqnarray}
where we shall choose $\alpha^2=2$. The physical coordinates
(\emph{solvable coordinates}) $\phi,\,\chi$, related to the four
dimensional fields, span half of this space (\emph{physical solvable
patch}) and are related to $X^\mu$ as follows:
$e^{-\phi}=X^1+X^0>0,\,e^{-\phi}\,\chi\propto X^2$, see Figure
\ref{fig1}. These coordinates correspond to choosing the coset
representative in the solvable subgroup of
${\rm SL}(2,\mathbb{R})$ defined by the Iwasawa decomposition and
will be referred to as the solvable parametrization:
\begin{equation}\label{cosetrep}
\mathbb{L}(\phi,\chi) = \rme^{\chi E} \rme^{\frac{\phi}{2} H} \,.
\end{equation}
The Lax operator in terms of the scalar fields and their
time-derivatives read:
\begin{eqnarray}
V(t)&=&\frac{1}{2}\,(\mathbb{L}^{-1}\dot{\mathbb{L}}+\eta\,\dot{\mathbb{L}}^T\,\mathbb{L}^{-T}\,\eta)=\nonumber\\&=&
\frac{1}{2}\, \left(\begin{matrix}\dot{\phi} &
e^{-\phi}\,\dot{\chi}\cr -e^{-\phi}\,\dot{\chi} & -
\dot{\phi}\end{matrix}\right)\,.\label{laxdot}
\end{eqnarray}
Time-like and light-like geodesics, see for instance
\cite{Breitenlohner:1987dg}, which arise from regular black holes in
four dimensions, are complete in the physical solvable patch.
Space-like geodesics on the other hand are not since they cross the
boundary $X^1+X^0=0$ at a finite value of the affine parameter $t$.
These solutions however are related to four dimensional black holes
with naked singularities. The origin $O$ of the solvable patch,
defined by $\phi=\chi=0$, can be mapped into a point $O^\prime$ in
the other half of the hyperboloid ($X^1+X^0<0$) by the compact
transformation:
\begin{eqnarray}
\mathcal{W}&=&\exp(\frac{\pi}{2}\,(E-E^T))=\left(\begin{matrix}0 &
1\cr -1 & 0\end{matrix}\right)\,.
\end{eqnarray}
This transformation, being in the coset, does not leave $\eta={\rm
}(+1,-1)$ invariant, but instead
$\mathcal{W}^{-1}\,\eta\,\mathcal{W}=-\eta$. The lower half of the
hyperboloid is still described by solvable coordinates
$\phi,\,\chi$, this time parametrizing the coset representative:
\begin{equation}
\mathbb{L}^\prime (\phi,\chi) = \rme^{\chi E} \rme^{\frac{\phi}{2}
H}\,\mathcal{W} \,.
\end{equation}
Let us choose the Lax operator at $t=0$ as follows:
\begin{equation}
V_0 = \left(
\begin{array}{cc}
 \frac{a}{\sqrt{2}} & \frac{k}{\sqrt{2}} \\
 -\frac{k}{\sqrt{2}} & -\frac{a}{\sqrt{2}}
\end{array}
\right)=\frac{a}{\sqrt{2}}\, \left(
\begin{array}{cc}
 1& -q \\
 q & 1
\end{array}
\right)\,,
\end{equation}
where $q=-\frac{k}{a}$. This is a diagonalizable initial condition
when $|q| \neq 1$. We consider the case where $|q|= 1$ separately
below. The final solution for the Lax operator $V_{\mathrm{sol}}(t)$
is given by
\begin{equation} V_{\mathrm{sol}}(t) = \left(\begin{array}{cc} V_{11}(t) & V_{12}(t) \\ - V_{12}(t) & -V_{11}(t) \end{array} \right)\,,
\end{equation}
where
\begin{align}
V_{11}(t) &=  \frac{a^2 f(t)-f(t) k^2-a g(t)
\sqrt{a^2-k^2}}{\sqrt{2} \left(a f(t)-g(t)
\sqrt{a^2-k^2}\right)} \,, \nonumber \\
 V_{12}(t) &= (k
\sqrt{a^2-k^2})/\nonumber\\&\sqrt{2} \Big(\text{cosh}\left[\sqrt{2}
t \sqrt{a^2-k^2}\right] \sqrt{a^2-k^2}\nonumber\\&-a\,
\text{sinh}\left[\sqrt{2} t \sqrt{a^2-k^2}\right]\Big) ,\end{align}
where we have defined
\begin{eqnarray}
f(t) & = & -1+\rme^{2 \sqrt{2} t \sqrt{a^2-k^2}} \,, \nonumber \\
g(t) & = & 1+\rme^{2 \sqrt{2} t \sqrt{a^2-k^2}} \,.
\end{eqnarray}
From the above Lax operator, the solutions for the scalar fields
can be found. They read, for $|q|<1$ as follows
\begin{eqnarray}
e^{-\phi(t)}&=&\frac{e^{-\phi_0}}{\sqrt{1-q^2}}\,\left[\sqrt{1-q^2}\,\cosh(x)-\sinh(x)\right]\,,\label{sol11}\\
\chi&=&\chi_0+\frac{e^{-\phi_0}}{q}-\frac{e^{-\phi_0}}{q}\,
\left[\frac{1+\frac{q^2}{2\,\sqrt{1-q^2}}\,\sinh(2\,x)}{1-\frac{q^2}{2\,(1-q^2)}\,(\cosh(2\,x)-1)}\right]\,,\label{sol12}\\
x&=&\sqrt{2\,(1-q^2)}\,a\,t\,,\nonumber
\end{eqnarray}
where $\phi_0$ and $\chi_0$ are the values of $\phi(t)$ and
$\chi(t)$ at $t=0$. These solutions are regular for
$x<\frac{1}{2}\,\log\left(\frac{1+\sqrt{1-q^2}}{1-\sqrt{1-q^2}}\right)$,
and thus are complete in the limit $a\,t\rightarrow -\infty$ . For
$|q|>1$ the solutions are obtained from the above expressions by
writing $x=i\,y=i\,\sqrt{2\,(q^2-1)}\,a\,t$:
\begin{eqnarray}
e^{-\phi(t)}&=&\frac{e^{-\phi_0}}{\sqrt{q^2-1}}\,\left[\sqrt{q^2-1}\,\cos(y)-\sin(y)\right]\,,\label{sol21}\\
\chi&=&\chi_0+\frac{e^{-\phi_0}}{q}-\frac{e^{-\phi_0}}{q}\,
\left[\frac{1+\frac{q^2}{2\,\sqrt{q^2-1}}\,\sin(2\,y)}{1+\frac{q^2}{2\,(q^2-1)}\,(\cos(2\,y)-1)}\right]\,,\label{sol22}
\end{eqnarray}
and are regular for $y_0<y<y_1$ where $y_1={\rm
Arctan}(\sqrt{q^2-1})>0$ and $y_0=y_1-\pi<0$ are the two points in
which $e^{-\phi}$ vanishes.
%
%
One can explicitly check that the above solutions
(\ref{sol11}),(\ref{sol12}) and (\ref{sol21}),(\ref{sol22}) do
indeed satisfy the geodesic equations. Let us also mention that
the norm squared of the geodesic is given by $2
\left(a^2-k^2\right)$. Since the eigenvalues $\lambda_{\pm}$ of
$V_0$ are given by $\lambda_{\pm}^2 = \frac{1}{2}\left(a^2 - k^2
\right)$, we see that geodesics with real eigenvalues have
positive norm squared, while the ones with imaginary eigenvalues
correspond to geodesics with negative norm squared.\par The
behavior of these solutions heavily depends on the values of $a$
and $k$. For instance, upon choosing $a = 2$ and $k = 1$, one
finds that the eigenvalues of the Lax operator are real.
The Lax operator is manifestly real and so are the solutions
(\ref{sol11}),(\ref{sol12}) for $\chi$ and $\phi$:
\begin{eqnarray}
e^{-\phi(t)} & =
&\frac{e^{-\phi_0}}{\sqrt{3}}\,\left[\sqrt{3}\,\cosh(\sqrt{6}\,t)-2\,\sinh(\sqrt{6}\,t)\right]\,,\nonumber\\
\chi(t) & = & \chi_0+2\,e^{\phi_0}-
\sqrt{3}\,e^{\phi_0}\left[\frac{4\,\sqrt{3}+\sinh(2\,\sqrt{6}\,t)}{7-\cosh(2\,\sqrt{6}\,t)}\right]\,.
\end{eqnarray}
In the limits $t\rightarrow \pm \infty$, the Lax operator $V(t)$
reduces to a diagonal matrix. One can also check that, in flowing
from $t=-\infty$ to $t=+\infty$, the eigenvalues remain constant.
In general, they will however get permuted on the diagonal during
the flow.\par For $a=1$ and $k=2$ on the other hand, one finds
that the Lax operator has purely imaginary eigenvalues $\lambda =
\rmi \sqrt{3/2}$ and $\bar{\lambda} = - \rmi \sqrt{3/2}$. The Lax
operator itself is still real however.
The solutions for $\phi$ and $\chi$ are given by  eqs.
(\ref{sol21}),(\ref{sol22}), which read:
\begin{eqnarray}
e^{-\phi(t)} & =
&\frac{1}{\sqrt{3}}\,e^{-\phi_0}\,\left[\sqrt{3}\,\cos(\sqrt{6}\,t)-\cos(\sqrt{6}\,t)\right]\,,\nonumber\\
\chi(t) & = & \chi_0+\frac{e^{\phi_0}}{2}-
\frac{\sqrt{3}}{2}\,e^{\phi_0}\,\left[\frac{\sqrt{3}+2\,\sin(2\,\sqrt{6}\,t)}{1+2\,\cos(2\,\sqrt{6}\,t)}\right]\,.
\end{eqnarray}
 The asymptotics for the Lax
operator are now very different from the case in which the
eigenvalues are real. The limits $t\rightarrow \pm \infty$ now no
longer lead to a diagonal Lax operator. Instead, these limits are
not well-defined due to the oscillating character of the functions
involved. What is physically relevant is the segment of the curve,
of finite length, contained in the physical solvable patch and
parametrized by $t\in ]t_0,\,t_1[$, corresponding to $y\in
]y_0,\,y_1[$, where $t_0<0$ and $t_1>0$ are the first zeros of
$e^{\phi}$ in the neighborhood of the origin $O$.\par Note that
for the Riemannian case, the Lax operator is a symmetric matrix
and the eigenvalues are always real. The Lax operator thus always
reduces to a diagonal matrix at $t=\pm \infty$. Indeed, this
phenomenon lies at the heart of the cosmic billiard phenomenon
(see e.g. \cite{Fre:2005bs, Fre':2007hd, Fre:2008zd}).\par In case
the Lax operator at $t=0$ is non-diagonalizable, namely $|q|=1$,
we can have $q=\varepsilon=\pm 1$.
 The initial value for the Lax operator is:
\begin{equation}
V_0 = \frac{a}{\sqrt{2}}\, \left( \begin{matrix}1 &
-\varepsilon\cr \varepsilon & 1\end{matrix}
\right)=\frac{a}{\sqrt{2}}\,n^\varepsilon \,,
\end{equation}
where $n^\varepsilon$ denotes two nilpotent elements such that, if
$J=\frac{1}{2}\,(E+E^T)=\frac{1}{2}\,\left(\begin{matrix}0 & 1 \cr
1& 0 \end{matrix}\right)$, we have: $[J,\,n^\varepsilon
]=\varepsilon\,n^\varepsilon$.
The solution can be found as the $|q|\rightarrow 1^-$ limit of
(\ref{sol11}),(\ref{sol12}) or, equivalently, as the
$|q|\rightarrow 1^+$ limit of (\ref{sol21}),(\ref{sol22}) and
reads:
\begin{eqnarray}
e^{-\phi(t)}&=&e^{-\phi_0}\,(1-\sqrt{2} \,a\,t)\,,\label{sol31}\\
\chi(t)&=&\chi_0-\varepsilon\,e^{\phi_0}\,\frac{\sqrt{2}\,a\,t}{1-\sqrt{2}\,a\,t}\,.\label{sol32}
\end{eqnarray}
This solution is regular as long as $\sqrt{2}\,a\,t<1$. The same
solution can be found by directly solving the Lax equation
$\dot{V}+[W,V]=0$. Indeed a nilpotent $V(t)$ in the coset can only
have the form: $V(t)=\frac{a(t)}{\sqrt{2}}\,n^\varepsilon$. In
this case we will have: $W=-\varepsilon\,\sqrt{2}\,a(t)\,J$. The Lax
equation is then equivalent to $\dot{a}(t)=-\sqrt{2}\,a(t)^2$,
which is solved by $a(t)=a/(1-\sqrt{2}\,a\,t)$, where we have
imposed $a(0)=a$. Substituting in $V(t)$ and using eq.
(\ref{laxdot}) one finds (\ref{sol31}) and (\ref{sol32}).\par This
example, analyzed in a certain detail, will be relevant to our
discussion of the generating geodesic of regular and small $D=4$
black holes, which will be done in Section \ref{ndic}.

\subsection{The $\SL(3,\mathbb{R})/\SO(2,1)$ example} \label{ssec:sl3}
\subsubsection{Solutions for diagonalizable initial conditions}

Let us now consider the example of geodesics on
$\SL(3,\mathbb{R})/\SO(2,1)$. We restrict ourselves here to
diagonalizable initial conditions. The Cartan generators in the
fundamental representation are given by
\begin{equation}
H_1 = \left(
\begin{array}{ccc}
 -\frac{1}{\sqrt{3}} & 0 & 0 \\
 0 & \frac{2}{\sqrt{3}} & 0 \\
 0 & 0 & -\frac{1}{\sqrt{3}}
\end{array}
\right) \,, \qquad H_2 = \left(
\begin{array}{ccc}
 -1 & 0 & 0 \\
 0 & 0 & 0 \\
 0 & 0 & 1
\end{array}
\right) \,,
\end{equation}
whereas the positive roots are given by
\begin{align}
E_{12} &= \left(
\begin{array}{ccc}
 0 & 1 & 0 \\
 0 & 0 & 0 \\
 0 & 0 & 0
\end{array}
\right),\qquad E_{23} = \left(
\begin{array}{ccc}
 0 & 0 & 0 \\
 0 & 0 & 1 \\
 0 & 0 & 0
\end{array}
\right),\qquad\nonumber\\
 E_{13} &= \left(
\begin{array}{ccc}
 0 & 0 & 1 \\
 0 & 0 & 0 \\
 0 & 0 & 0
\end{array}
\right)\,.
\end{align}
The subspaces $\mathbb{H}$ and $\mathbb{K}$ that determine the
Cartan decomposition are then explicitly given by
\begin{align}
\mathbb{H} & = \mathrm{Span} \left\{\left(E_{12} +
E_{12}^T\right),\left(E_{23} + E_{23}^T\right),\left(E_{13} - E_{13}^T\right)  \right\} \nonumber\\
\mathbb{K} & =  \mathrm{Span} \left\{H_1, H_2, \left(E_{12} -
E_{12}^T\right),\left(E_{23} - E_{23}^T\right),\left(E_{13} +
E_{13}^T\right) \right\} \nonumber \,.
\end{align}

We will define the coset representative as follows
\begin{equation}
\mathbb{L} = \rme^{\chi_1(t) E_{12}} \rme^{\chi_0(t) E_{23}}
\rme^{\chi_2(t) E_{13}} \rme^{\frac{1}{2} \phi_1(t) H_1 +
\frac{1}{2} \phi_2(t) H_2} \,.
\end{equation}
Let us illustrate the behaviour of the solutions with
diagonalizable initial conditions by giving two specific examples.

\paragraph{All eigenvalues are real:}

The first initial condition we take, is characterized by the
following Lax operator at $t=0$:
\begin{equation}
V_0 = \left(
\begin{array}{ccc}
 -1 & 0 & 0 \\
 0 & \frac{1}{2} (1+3\ \text{cosh}[\pi ]) & -\frac{3\ \text{sinh}[\pi ]}{2} \\
 0 & \frac{3\ \text{sinh}[\pi ]}{2} & \frac{1}{2} (1-3\ \text{cosh}[\pi ])
\end{array}
\right)\,.
\end{equation}
The eigenvalues for the Lax operator are all real and given by
$-1$, $2$ and $-1$. Running the Lax algorithm, the solutions for
the scalars are easily found to be:
\begin{eqnarray}
\phi_1(t) & = & C_1+\frac{1}{2} \sqrt{3}
(4 t-\text{log}[-1+\rme^{6 t} (-1+\text{cosh}[\pi ])\nonumber\\&-& \text{cosh}[\pi ]])  \nonumber \,, \\
\phi_2(t) & = &  C_2+\frac{1}{2} \text{log}
\left[-1+\rme^{6 t} (-1+\text{cosh}[\pi ])-\text{cosh}[\pi ]\right] \nonumber \,, \\
\chi_0(t) & = & C_4+\frac{\rme^{\frac{1}{2} \left(\sqrt{3}
C_1-C_2\right)} \text{coth}
\left[\frac{\pi }{2}\right]}{1-\rme^{6 t} (-1+\text{cosh}[\pi ])+\text{cosh}[\pi ]} \nonumber \,, \\
\chi_1(t) & = &  C_3  \nonumber \,, \\
\chi_2(t) & = &  C_5  \nonumber \,,
\end{eqnarray}
where $C_1,\cdots,C_5$ are arbitrary integration constants. The
behaviour of the Lax operator is very similar to the
$\SL(2,\mathbb{R})$-example with real eigenvalues. Again, in flowing
from $t=-\infty$ to $t=+ \infty$ the eigenvalues remain constant and
get permuted on the diagonal. This example again corresponds to a
geodesic with a positive length squared.

\paragraph{Some eigenvalues are complex:}

The second initial condition we take, is characterized by the Lax
operator at $t=0$:
\begin{equation}
V_0 = \left(
\begin{array}{ccc}
 -1 & 0 & 0 \\
 0 & 1 & 1 \\
 0 & -1 & 0
\end{array}
\right)\,.
\end{equation}
In this case, one eigenvalue is real and given by $-1$, while the
other two eigenvalues are complex and given by $\frac{1}{2}
\left(1+\rmi \sqrt{3}\right)$ and $\frac{1}{2} \left(1-\rmi
\sqrt{3}\right)$ \footnote{Note that since $V_0$ is real, the
complex eigenvalues always come with their complex conjugate. For
the $\SL(3,\mathbb{R})$ example, one will thus always have one
real eigenvalue.}. Upon applying the Lax algorithm, one finds the
following solutions for the scalar fields:
\begin{eqnarray}
\phi_1(t) & = & C_1-\frac{1}{2} \sqrt{3} \Big(-t+\text{log}\big[-3\
\text{cos} \big[\sqrt{3} t\big]+\nonumber\\&&\sqrt{3}\ \text{sin}
\big[\sqrt{3} t\big]\big]\Big) \nonumber \,, \\
\phi_2(t) & = &  C_2+\frac{1}{2} \Big(3 t+\text{log}\big[-3
\text{cos}\big[\sqrt{3} t\big]+\nonumber\\&&\sqrt{3}\ \text{sin}
\big[\sqrt{3} t\big]\big]\Big)\nonumber \,, \\
\chi_0(t) & = & C_4+\frac{2 \rme^{\frac{1}{2} \left(\sqrt{3}
C_1-C_2\right)} \text{sin} \left[\sqrt{3} t\right]}{3
\left(-\sqrt{3}\ \text{cos}
\left[\sqrt{3} t\right]+\text{sin}\left[\sqrt{3} t\right]\right)} \nonumber \,, \\
\chi_1(t) & = &  C_3  \nonumber \,, \\
\chi_2(t) & = &  C_5  \nonumber \,,
\end{eqnarray}
As in the $\SL(2,\mathbb{R})$ case with complex eigenvalues, the
limits $\lim_{t\rightarrow \pm \infty} V(t)$ are not well-defined
and hence do not lead to a diagonal Lax operator, due to the
oscillating behaviour of the functions involved in the solutions.
In this case, the corresponding geodesic is null-like. A geodesic
with negative norm squared can be obtained by applying the Lax
algorithm with initial condition
\begin{equation}
V_0 = \left(
\begin{array}{ccc}
 -1 & 0 & 0 \\
 0 & 1 & 2 \\
 0 & -2 & 0
\end{array}
\right)\,,
\end{equation}
which again has one real eigenvalue and two complex eigenvalues,
that are each others complex conjugates.
\subsubsection{Regular solutions for nilpotent initial conditions} \label{sssec:regsl3}
After running the Lax algorithm for the initial condition
\begin{equation} \label{V0}
V_0 = \left(\begin{array}{ccc} 0 & 0 & 1 \\ 0 & 0 & a \\ 1 & -a &
0 \end{array} \right)\,,
\end{equation}
we obtain the following solutions for the scalars (for $1-a^2 >
0$):
\begin{align} \label{sols}
&\phi_1(t)  =  C_1-\frac{1}{2} \sqrt{3}\  \text{log}
\left[\text{cosh}\left[2 \sqrt{1-a^2}
t\right]\right]+\frac{\sqrt{3}}{2}\ \text{log}
\big[\nonumber\\&-a^2+\text{cosh}\big[2 \sqrt{1-a^2} t\big]\big] \,, \nonumber \\
&\phi_2(t)  =  C_2 +\frac{1}{2} \Big(\text{log}
\big[\text{cosh}\big[2 \sqrt{1-a^2} t\big]\big]+\text{log}
\big[-a^2\nonumber\\&+\text{cosh}\big[2 \sqrt{1-a^2} t\big]\big]\Big) \,, \nonumber \\
&\chi_0(t) =  C_4 +a \ \rme^{\frac{1}{2} \left(\sqrt{3}
C_1-C_2\right)}\ \text{tanh}
\left[2 \sqrt{1-a^2} t\right]\,, \nonumber \\
&\chi_1(t)  =  \frac{\rme^{\frac{1}{2} \big(-\sqrt{3}
C_1-C_2\big)}}{a^2-\text{cosh}\big[2 \sqrt{1-a^2} t\big]}
 \times\Big(a+\nonumber\\&a^2 \rme^{\frac{1}{2} \big(\sqrt{3} C_1+C_2\big)} C_3 - \rme^{\frac{1}{2}
  \big(\sqrt{3} C_1+C_2\big)} C_3 \ \text{cosh}
  \big[2 \sqrt{1-a^2} t\big]\big) \,, \nonumber \\
&\chi_2(t) =  C_5+\nonumber\\&\frac{\rme^{-\frac{\sqrt{3}
C_1}{2}-C_2} \left(a\ \rme^{C_2/2}\ C_4+\rme^{\frac{\sqrt{3}
C_1}{2}} \text{sinh}\left[2 \sqrt{1-a^2}
t\right]\right)}{-a^2+\text{cosh}\left[2 \sqrt{1-a^2} t\right]}
\,.
\end{align}
In the limit $a\rightarrow 1$, the initial condition (\ref{V0})
becomes nilpotent and the solutions (\ref{sols}) become singular.
One can however renormalize the constants $C_1$ and $C_2$ in such
a way that the renormalized solutions are regular in the limit
$a\rightarrow 1$ and are still valid solutions of the geodesic
equations in this limit.

How this renormalization should be performed can be easily seen by
looking at the series expansion of the solutions around $a=1$. For
$\phi_1(t)$ and $\phi_2(t)$, we get
\begin{eqnarray}
\phi_1(t) & = &  \left(C_1+\frac{\sqrt{3}}{2}
\text{log}[-1+a]+\frac{\sqrt{3}}{2} \text{log} \left[-2-4
t^2\right]\right)+\nonumber\\& &\frac{\sqrt{3}
\left(3+30 t^2+40 t^4\right) (a-1)}{12+24 t^2} +O[a-1]^{3/2},\\
\phi_2(t) & = &
\Big(C_2+\frac{1}{2}\text{log}[-1+a]+\frac{1}{2}\text{log}\big[-2-4
t^2\big]\Big)+\nonumber\\& &\frac{\big(3-18 t^2-56 t^4\big)
(a-1)}{12 \big(1+2 t^2\big)}+O[a-1]^{3/2}  \,. \nonumber
\end{eqnarray}
(Let us ignore for the moment the fact that some terms seem
complex instead of real. We shall deal with that later.) In both
cases, the infinity in taking the limit $a\rightarrow 1$ comes
from the second term in the expansion, while all other terms are
completely regular.

The solutions for $\phi_1(t)$ and $\phi_2(t)$ can thus be made
regular in this limit, by renormalizing (redefining) $C_1$ and
$C_2$ as
\begin{eqnarray}
C_1 & \rightarrow &  \frac{1}{2} \sqrt{3}\ \text{log}\left[-\frac{p_1}{-1+a}\right] \,, \nonumber \\
C_2 & \rightarrow & \frac{1}{2}\
\text{log}\left[-\frac{p_2}{-1+a}\right] \,.
\end{eqnarray}
Note that we will take the new integration constants $p_1$ and
$p_2$ to be positive. From the series expansion above, it is then
immediate that the solutions for $\phi_1(t)$ and $\phi_2(t)$
become regular in the limit $a\rightarrow 1$:
\begin{eqnarray}
\phi_1(t) & = & \frac{1}{2} \sqrt{3} \left(\text{log}\left[p_1(2+4 t^2)\right]\right) \,, \nonumber \\
\phi_2(t) & = &  \frac{1}{2} \left(\text{log}\left[p_2(2+4
t^2)\right]\right) \,.
\end{eqnarray}
It's an amusing fact that the constants $C_1$ and $C_2$ appear in
the solutions for $\chi_0(t)$, $\chi_1(t)$ and $\chi_2(t)$ in such
a way as to render them regular as well in the limit $a\rightarrow
1$ after performing the renormalization:
\begin{eqnarray}
\chi_0(t) & = &  \frac{-2 \sqrt{2}\ (p_1 p_2)^{3/4} t}{p_2}+C_4 \,, \nonumber \\
\chi_1(t) & = &  -\frac{1}{2 \left(p_1^3 p_2\right)^{1/4} \left(1+2 t^2\right)}+C_3 \,, \nonumber \\
\chi_2(t) & = & -\frac{2 \sqrt{2} p_1 \sqrt{p_2} t-\left(p_1
p_2^3\right)^{1/4} C_4}{2 p_1 p_2+4 \text{p1} p_2 t^2}+C_5 \,.
\end{eqnarray}
Note that the above solutions are real for $p_1, p_2 >0$. One can
also check that they obey the geodesic equations\footnote{We have
learned that P. Fr\'e and A. Sorin have independently obtained a
similar result.}.
\section{Relation to the black hole generating geodesic}\label{ndic}
As already emphasized, the algorithm
discussed in the present note is well defined only for solutions
corresponding to diagonalizable initial conditions. Although these
solutions may yield, in certain limits, geodesics with
non-diagonalizable initial data, it is not proven that the most
general solution of the latter type can be obtained in this way.
However the generating geodesic  of regular $D=4$ black holes
(including those with vanishing horizon area, i.e. \emph{small}
black holes), defined as the simplest solution capturing all the
$G$-invariant properties of the most general one, can be obtained
as a singular limit of solutions with diagonalizable $Q$. This
generating geodesic was described in \cite{Bergshoeff:2008be} as a
geodesic within a simple characteristic submanifold of $G/H$
consisting of a product of $dS_2$ spaces times $\mathbb{R}_+$
factors. This analysis also applies to extremal solutions, since
one can show that the Noether charge matrices of light-like
geodesics in this simple submanifold have representatives in all
the nilpotent orbits corresponding to the regular extremal $D=4$
solutions (including small black holes). Let us first recall the
construction made in \cite{Bergshoeff:2008be} of the generating
geodesic of regular $D=4$ black holes. Then we shall further
elaborate on it and work out the explicit transformation mapping
the BPS and non-BPS extremal solutions. This transformation
belongs to the complexification $G^\mathbb{C}$ of $G$ but not to
$G$ and its action on the generating geodesics is quite easy to
characterize, having the latter very simple form. For the sake of
simplicity we shall restrict our analysis to the maximally
supersymmetric supergravity and to the $\mathcal{N}=2$ STU
model.\par We need to first briefly recall the relation between
$D=4$ static, single center, asymptotically flat black holes and
geodesics on $G/H$. Let us start from a $D=4$ extended
supergravity with a symmetric scalar manifold of the form
$G_4/H_4$, spanned by $n_S$ scalars $\phi^r$ and describing $n_V$
vector fields $A^\Lambda_\mu$. Being $G_4/H_4$ Riemannian, it is
globally described by the solvable group
$\mathcal{S}_4=\exp(Solv_4)$ defined by the Iwasawa decomposition
of $G_4$ with respect to $H_4$ and parametrized by $\phi^r$:
$Solv_4=\{\phi^r s_r\}$. The ansatze for the four dimensional
metric and the symplectic vector $\mathbb{F}^M_{\mu\nu}$
consisting of the electric field strengths $F^\Lambda_{\mu\nu}$
and their magnetic duals $G_{\Lambda\,\mu\nu}\propto \,{}^\star
\delta \mathcal{L}_4/\delta F^\Lambda$ read:
\begin{eqnarray}
\rmd s^2&=&-e^{2\,U}\,\rmd
t^2+e^{-2\,U}\,\left[\frac{c^4}{\sinh^4(c\,\tau)}\,\rmd\tau^2+
\frac{c^2}{\sinh^2(c\,\tau)}\,(\rmd \theta^2+\sin(\theta)\,\rmd \varphi^2)\right]\,,\nonumber\\
\mathbb{F}&=&\left(\begin{matrix}F^\Lambda_{\mu\nu}\cr
G_{\Lambda\,\mu\nu}
\end{matrix}\right)\,\frac{dx^\mu\wedge
dx^\nu}{2}=-e^{2\,U}\mathbb{C}\cdot\mathcal{M}_4(\phi^r)\cdot\Gamma\,\rmd
t\wedge
\rmd\tau+\Gamma\,\sin(\theta)\,\rmd\theta\wedge
\rmd\varphi\,,
\end{eqnarray}
where $c$ is the \emph{extremality parameter}, $\mathbb{C}^{MN}$
denotes the $2\,n_V\times 2\,n_V$ symplectic invariant matrix,
$\mathcal{M}_4(\phi^r)\equiv-\mathbb{L}_4\,\mathbb{L}_4^T<0$,
$\mathbb{L}_4\in \mathcal{S}_4$ being the $G_4/H_4$ coset
representative in the symplectic $2\,n_V$-dimensional
representation, and $\Gamma^M\equiv(p^\Lambda,\,q_\Lambda)$ the
quantized magnetic and electric charges. Let us stress that in the
present section the radial variable $\tau$ plays the role of the
variable $t$ of the previous sections, not to be confused with the
time coordinate $t$. The global symmetry group of the $D=4$ theory
is $G_4$ which acts simultaneously as an isometry on $\phi^r$ and
by means of linear symplectic electric-magnetic transformations on
$\mathbb{F}^M$. The charge vector $\Gamma^M$ therefore transform
in a symplectic representation ${\bf R}$ of $G_4$. The radial
variable $\tau$ is by definition negative. The horizon is located
at $\tau\rightarrow -\infty$ while radial infinity corresponds to
$\tau\rightarrow 0^-$.\par Upon reduction along the time direction
to $D=3$ and dualization of the $D=3$ vector fields into scalars,
we end up with $D=3$ gravity coupled to a sigma model whose
$2+2\,n_V+n_S$ scalars $\phi^I$ consist in $U$ and the dual $a$ to
the Kaluza-Klein vector, coming from the metric, the $D=4$ scalars
$\phi^r$ and $2\,n_V$ scalar fields
$\mathcal{Z}^M=(\zeta^\Lambda,\,\tilde{\zeta}_\Lambda)$
originating from the vector fields and satisfying the relation:
$\frac{\rmd}{\rmd \tau}\mathcal{Z}^M=\mathbb{F}^M_{0\tau}$. The
sigma model metric reads:
\begin{eqnarray}
\rmd {\bf s}^2&=&G_{IJ}\,\rmd \phi^I\,\rmd \phi^J=2\,(\rmd
U)^2+G_{rs}\,\rmd \phi^r\,\rmd
\phi^s+\frac{1}{2}\,e^{-4 \,U}\,\omega^2+e^{-2
\,U}\,\rmd \mathcal{Z}^T\cdot\mathcal{M}_4\cdot\rmd \mathcal{Z}\,,
\end{eqnarray}
where $\omega\equiv \rmd a+\mathcal{Z}^T\,\mathbb{C}\,d\mathcal{Z}$.
The signature of $G_{IJ}$ clearly has $2\,n_V$ minus signs
corresponding to the $d\mathcal{Z}$ directions. The global symmetry
group $G$ of the $D=3$ Euclidean theory contains the product
$\SL(2,\mathbb{R})_E\times G_4$, where $\SL(2,\mathbb{R})_E$,
generated by $\mathfrak{sl}(2,\mathbb{R})_E\equiv {\rm
Span}(H_{\beta_0},\, E_{\beta_0},\, E_{-\beta_0}$), see footnote 4,
is the Ehlers group acting transitively on $U,\,a$. Similarly $H$
contains $\U(1)_E\times H_4$, where $\U(1)_E$ is the compact
subgroup of $\SL(2,\mathbb{R})_E$, generated by $E_{\beta_0}-
E_{-\beta_0}$. The charge vector $\Gamma^M$ transforms in a
symplectic representation of $\U(1)_E\times H_4$ which we shall
still denote, with an abuse of notation,  by ${\bf R}$. The solvable
group which describes the \emph{physical patch} of $G/H$ is defined
by the Iwasawa decomposition of $G$ with respect to its maximal
compact subgroup (i.e. the compact form of the complexification of
$H$) and its generating solvable Lie algebra $Solv={\rm Span}(s_I)$
is parametrized by $\phi^I$. The algebra $Solv$ decomposes with
respect to $Solv_4$, parametrized by $\phi^r$, as follows:
\begin{eqnarray}
Solv&=&{\rm Span}(H_{\beta_0},\,E_{\beta_0})\oplus Solv_4\oplus {\bf
R}_{+1}\,,
\end{eqnarray}
where the space ${\bf R}_{+1}\equiv {\rm Span}(s_M)$ is parametrized
by the scalars $\mathcal{Z}^M$ and the grading refers to
$H_{\beta_0}$. Together with the grading $+2$ generator
$E_{\beta_0}$, the nilpotent generators $s_M$ close a Heisenberg
algebra: $[s_M,\,s_N]=\frac{1}{2}\,\mathbb{C}_{MN}\,E_{\beta_0}$.
The coset representative of $G/H$ in the physical patch is thus
defined as follows: $\mathbb{L}=\exp(a \,E_{\beta_0})\,
\exp(\sqrt{2}\,\mathcal{Z}^M\,s_M)\,\exp(\phi^r\,s_r)\,\exp(U\,H_{\beta_0})\in
\exp(Solv)$. \par We can now split the spaces $\mathbb{K}$ and
$\mathbb{H}$ in the Cartan decomposition of $\mathbb{G}$ with
respect to $\mathbb{H}$, as follows:
\begin{eqnarray}
\mathbb{K}&=&\mathbb{K}_2\oplus \mathbb{K}_4\oplus
\mathbb{K}^{(R)}\,;\,\, \mathbb{H}=\mathfrak{u}(1)_E\oplus
\mathbb{H}_4\oplus \mathbb{H}^{(R)}\,,\nonumber
\end{eqnarray}
where $\mathbb{K}_2={\rm
Span}(H_{\beta_0},\,E_{\beta_0}+E_{-\beta_0})$ is the compact
subspace of $\mathfrak{sl}(2,\mathbb{R})_E$ while $\mathbb{K}_4$
is the non-compact space defined by the Cartan decomposition of
the algebra $\mathbb{G}_4$ generating $G_4$ with respect to its
maximal compact subalgebra $\mathbb{H}_4$. The spaces
$\mathbb{K}^{(R)}\equiv {\rm Span}(k^M)$ and
$\mathbb{H}^{(R)}\equiv {\rm Span}(s_M-s_M^\dagger)$,
$k^M=s_M+s_M^\dagger$, transform in the representation ${\bf R}$
with respect to $\U(1)_E\times H_4$. The former consists of
compact matrices and define the negative signature directions on
$G/H$. The latter consists of non-compact matrices and generates
the Riemannian coset $H/[\U(1)_E\times H_4]$. We can understand
the $D=4$ interpretation of the constants of motion encoded in the
matrix $Q$, by restricting, for the sake of simplicity, to
geodesics originating at radial infinity, from the origin,
$\phi^I=0$, of the physical patch:
$\mathbb{L}[\phi^I(\tau=0)]={\bf 1}$. In this case, from eq.
(\ref{QV}), we can express the corresponding Noether matrix
$Q=Q_0$ in terms of the Lax operator at infinity: $Q_0=2\,V_0^T\in
\mathbb{K}$, where $V_0=V(\tau=0)$. The components of $Q_0$ along
$H_{\beta_0}$ and $E_{\beta_0}+E_{-\beta_0}$ are the ADM mass and
the NUT charge, the components along $\mathbb{K}_4$ are the scalar
charges $\Sigma_r$ while its projection $Q^{(R)}$ along
$\mathbb{K}^{(R)}$ are the electric and magnetic charges:
\begin{eqnarray}Q^{(R)}&\equiv &Q_0\cap
\mathbb{K}^{(R)}=-\sqrt{2}\,\Gamma^M\mathbb{C}_{MN}\,
k^M=-\sqrt{2}\,(p^\Lambda\,k_\Lambda-q_\Lambda\,k^\Lambda)\,.\label{qr}\end{eqnarray}
What is the minimal set of generators of $\mathbb{K}^{(R)}$ or of
$\mathbb{H}^{(R)}$ along which a generic element of these spaces
can be rotated by means of a $\U(1)_E\times H_4$ transformation?
This minimal set has dimension $p\equiv {\rm
rank}(H/[\U(1)_E\times H_4])$ and defined the maximal set of
commuting elements $\{J_k\}$ of $\mathbb{H}^{(R)}$, or, the
maximal number of commuting elements $\{k_k\}$ of
$\mathbb{K}^{(R)}$, $k=0,\dots, p-1$. More specifically it defines
the dimension of the \emph{normal form} of the representation
${\bf R}$ of $\U(1)_E\times H_4$. In other words the minimal
number of electric and magnetic charges which characterizes the
most general geodesic modulo $G$-transformations (i.e. the
\emph{generating geodesic}) is $p$. Consider for instance the
maximal supergravity in $D=4$. We have $G_4=\E_{7(7)}$,
$H_4=\SU(8)$. Upon reduction to $D=3$ the global symmetry gets
enhanced to $G=\E_{8(8)}$. In this case $H=\SO^*(16)$ and the
$D=3$ scalar manifold is therefore $\E_{8(8)}/\SO^*(16)$. The
quantized charges transform in the representation ${\bf 56}$ of
$\E_{7(7)}$, while the central charge matrix $Z_{AB}$,
$A,B=1,\dots, 8$, and its conjugate $\overline{{Z}}^{AB}$ belong
to the  ${\rm 28}_{+1}+\overline{{\rm 28}}_{-1}$ of $\U(1)_E\times
H_4=\U(8)$. Since we are considering geodesics stemming from the
origin $\phi^I(0)=0$, at radial infinity the central charges
$(\bar{Z}^{AB},\,Z_{AB})_{\vert\tau=0}$ and the quantized charges
$\Gamma^M$ are related by a basis transformation. The
corresponding component of $Q_0$ reads: $Q^{(R)}\propto
(\bar{Z}^{AB}\,k_{AB}-Z_{AB}\,k^{AB})_{\vert\tau=0}$. It is known
that a complex $8\times 8 $ matrix can always be skew-diagonalized
by means of a $\U(8)$ transformation. If we denote by $\rho_k$ the
four real skew-eigenvalues of $Z_{AB}^{(0)}=Z_{AB}(\tau=0)$, we
can therefore, through a suitable $\U(8)$ conjugation, bring
$Q^{(R)}$ to its \emph{normal form}: $Q^{(R)}\rightarrow
\rho_k\,k_k$. In this case the four compact generators $k_k$
generate the maximal abelian subalgebra of the $56$-dimensional
space $\mathbb{K}^{(R)}$ and $p={\rm rank}(\SO^*(16)/\U(8))=4$
coincides indeed with the dimension of the normal form. In general
we can choose the generators $(k_k)$ in $\mathbb{K}^{(R)}$ and
$(J_k)$ in $\mathbb{H}^{(R)}$ so that, together with
$H_k\equiv[J_k,\,k_k]$, generate an $\SL(2,\mathbb{R})^p$ subgroup
of $G$. In particular $(H_k,\,k_k)$ generate a submanifold
$\left(\frac{\SL(2,\mathbb{R})}{\SO(2)}\right)^p$ of $G/H$ where,
as we shall see, we can find the generating geodesic of extremal
(regular) black holes in $D=4$. This kind of solutions are
generated by a nilpotent $V_0$ (i.e. $Q$) and moreover can be
obtained as a singular limit of solutions with diagonalizable
$V_0$. In \cite{Bergshoeff:2008be} it was shown that solutions
with diagonalizable $V_0$ can be brought, by means of a $G$
transformation, to lie in the  following submanifold of $G/H$:
\begin{eqnarray}
\mathcal{M}_N&=&\left(\frac{\SL(2,\mathbb{R})}{\SO(1,1)}\right)^p\times\SO(1,1)^{r-p}=
\exp(\mathbb{K}_N)\,,\label{mn}
\end{eqnarray}
where the first $p$ factors are generated by $(H_k,\,k_k)$. A
geodesic in $G/H$, defined by a diagonalizable $V_0$ and an initial
point $p_0=(\phi^I_0)=(\phi^I(0))\in G/H$, can then be mapped into a
geodesic on $\mathcal{M}_N$ by first mapping $p_0$ into a point
$p^\prime_0$ of $\mathcal{M}_N$, by means of a transformation in
$G/H$, and then using the action of the isotropy group $H$ of
$p^\prime_0$ to rotate $V_0$ into the tangent space to
$\mathcal{M}_N$ at $p^\prime_0$. The manifold $\mathcal{M}_N$ will
be referred to as the \emph{normal form} of $G/H$ for solutions with
diagonalizable Lax operator\footnote{Note that in
\cite{Bergshoeff:2008be} the normal form of
$\frac{\GL(p,q)}{\SO(p,q)}$ has been presented in full generality.}.
%
 \par Since the $p$
$\SL(2,\mathbb{R})$ groups in (\ref{mn}) are defined
\cite{Bergshoeff:2008be} by a maximal set of $p$-commuting
generators out of those corresponding to the $D=4$ electric-magnetic
charges plus a set of 4 corresponding  Cartan generators, a geodesic
of $\mathcal{M}_N$ will describe a four-dimensional dilatonic
solution coupled to $p$ vector fields. The dilatons parametrizing
the $\SO(1,1)^{r-p}$ factor in (\ref{mn}) are decoupled from the
charges.\par
%
These considerations do not apply to solutions with
non-diagonalizable initial conditions, such as those originating
from extremal, non-rotating, black holes in $D=4$, for which ${\rm
Tr}(V_0^2)=0$ and moreover $V_0$ is nilpotent. In these cases $V_0$
is classified within nilpotent orbits with respect to $H$. In
\cite{Bossard:2009at} the nilpotency condition was obtained for a
$V_0$ in the fundamental representation of $G$, as the ${\rm
Tr}(V_0^2)\rightarrow 0$ limit of a general relation satisfied by
the non-extremal solutions and reads:
\begin{eqnarray}
V_0^3&=&0\,\,\,\,\,\,\,(G\neq
\E_{8(8)},\,\E_{8(-24)})\,\,;\,\,\,\,\,\nonumber\\V_0^5&=&0\,\,\,\,\,\,\,(G=
\E_{8(8)}\,\,\mbox{or}\,\,\,\E_{8(-24)})\,.\label{nc}
\end{eqnarray}
It turns out that nilpotent elements of the subspace
$\mathbb{K}_N$ of $\mathbb{K}$ have, for different choices of
their parameters, representatives in all the relevant orbits
defined by (\ref{nc}).\par In what follows we shall discuss the
generating geodesic of regular extremal black holes in the
maximally supersymmetric theory reviewing and extending the
analysis in \cite{Bergshoeff:2008be}. The manifold $\mathcal{M}_N$
associated with this model is the same as the one associated with
the $D=3$ model with scalar manifold:
\begin{eqnarray}
\frac{G^\prime}{H^\prime}&=&\frac{\SO(4,4)}{\SO(2,2)\times
\SO(2,2)}\times\frac{\SO(4,4)}{\SO(4)\times
\SO(4)}\,,\label{stuman}
\end{eqnarray}
which originates from the $\mathcal{N}=2, D=4$ $STU$-model coupled
to four hypermultiplets. This is consistent with the known
property that the $STU$-truncation of the $\mathcal{N}=8$ model
describes the generating (seed) solution
\cite{Gimon:2007mh,Andrianopoli:1997wi,Bertolini:2000ei} of
extremal, black holes in the maximal theory. The second factor on
the right-hand side of  (\ref{stuman}) is parametrized by
hyperscalars. A generator $V_0$ in the $[\mathfrak{so}(4,4)\ominus
(\mathfrak{so}(2,2)\oplus\mathfrak{so}(2,2))]$ space, generating
the first factor, transforms in the ${\bf (2,2,2,2)}$
representation of the group $\SO(2,2)\times
\SO(2,2)\equiv\SL(2,\mathbb{R})^4$ at the denominator. It can be
written in components as
\begin{eqnarray}
V_0&=&\sum V_0^{A_1,A_2,A_3,A}\vert
A_1,A_2,A_3,A\rangle\,,\label{4qbit}\end{eqnarray}
 where $\vert A_1,A_2,A_3,A\rangle$, $A, \,A_i=1,2$, is a basis of matrices; $A$
also labels the supersymmetry parameter $\epsilon_A$ and is the
doublet index of the pseudo-quaternionic structure group ${\rm
SL}(2,\mathbb{R})_0\subset \SL(2,\mathbb{R})^4$ \footnote{The
supersymmetry variation of the eight fermionic fields in the
quarter-maximal theory $\lambda^{A_1,A_2,A_3}$ reads: $\delta
\lambda^{A_1,A_2,A_3}=V_0^{A_1,A_2,A_3,A}\,\epsilon_A$.}. The
matrix $V_0$ in eq. (\ref{4qbit}) can also be seen as an element of
a \emph{4 q-bit} system. Consider regular static black hole
solutions. The BPS one corresponds to a \emph{factorized}
nilpotent initial condition:
$V_0^{A_1,A_2,A_3,A}=C^{A_1,A_2,A_3}\,v^A$ and $V_0^3=0$ with
$V_0$ in the fundamental of $\SO(4,4)$. In
\cite{Bergshoeff:2008be} it was shown that the $\mathcal{N}=2$
non-BPS extremal  solution with positive quartic invariant (and
vanishing central charge at the horizon), correspond to a
different factorization:
$V_0^{A_1,A_2,A_3,A}=C^{A_i,A_j,A}\,v^{A_k}$, $i,j,k=1,2,3$. These
solutions, together with the $\mathcal{N}=2$ BPS one correspond to
$\mathcal{N}=8$ BPS black hole solutions. Regular solutions with
non-factorized initial data $V_0$ define non-BPS black holes with
negative quartic invariant. These correspond to the case in which
$V_0$ is an \emph{entangled} state in the 4 q-bit system.\par Let
us now consider the solutions to the smaller model based on the
manifold $\mathcal{M}_N$. Being this manifold the product of $p=4$
$dS_2$ factors times $r-p$ ${\rm O}(1,1)$ factors, a geodesic on
it is the product of the $p$ geodesics within each $dS_2$ factor,
which were discussed in Section \ref{ssec:sl2}, times $r-p$
geodesics in each ${\rm O}(1,1)$ factor. The dilatons in the
$\SO(1,1)^{r-p}=\SO(1,1)^4$ factor of $\mathcal{M}_N$, from the
$\mathcal{N}=2$ point of view, are hyperscalars in the
$\mathcal{N}=2$ theory and therefore will not be relevant to our
discussion of black holes.\par Let us consider extremal geodesics
on $\mathcal{M}_N$, which defined by a nilpotent $V_0$ in
$\mathbb{K}_N \bigcap [\mathfrak{so}(4,4)\ominus
(\mathfrak{so}(2,2)\oplus\mathfrak{so}(2,2))]$, in the fundamental
eight-dimensional representation of $\SO(4,4)$. Being nilpotent,
$V_0$ can only belong to the generators of the
$\left(\frac{\SL(2,\mathbb{R})}{\SO(2)}\right)^p$ factors.
\par Recall
that the generators $(k_k,H_k)$ of the
$\left(\frac{\SL(2,\mathbb{R})}{\SO(2)}\right)^p$ manifold were
constructed out of the normal form of the electric and magnetic
charges in $D=4$. If the four dimensional STU model originates
from reduction of a $D=5$ theory, its scalar manifold is described
by the \emph{special coordinate frame} and normal forms are the
$D0-D4-D4-D4$ (anti-) brane charges $q_0,\,p^1,\,p^2,\,p^3$ or the
$D6-D2-D2-D2$ (anti-) brane charges $p^0,\,q_1,\,q_2,\,q_3$. Each
subspace $[\mathfrak{sl}(2,\mathbb{R})\ominus
\mathfrak{so}(1,1)]_k$ within $\mathbb{K}_N$ is generated by the
nilpotent shift matrices $n_k^{\pm}=H_k\pm
k_k=\left(\begin{matrix}1 & \mp 1\cr \pm 1 &
-1\end{matrix}\right)$ such that $[J_k,n_k^\pm]=\pm n_k^\pm$,
where $J_k=\frac{1}{2}\,\left(\begin{matrix}0 & 1\cr  1 &
0\end{matrix}\right)$\footnote{Each $2\times 2$ matrix, for
different values of  $k$, should be thought of as acting on a
different space.}. A nilpotent $V_0$ in $\mathbb{K}_N$ will have
the following general form:
\begin{eqnarray}
 V_0 &\in&  \oplus_{k=1}^4[\mathfrak{sl}(2,\mathbb{R})\ominus
\mathfrak{so}(1,1)]_k\,\,,\,\,\, V_0^n=0\,\,\,\Rightarrow
\,\,\,\,\,\,V_0=\frac{a_0}{\sqrt{2}}\,
n^{-\varepsilon_0}+\sum_{i=1}^3 \frac{a_i}{\sqrt{2}}\,
n^{\varepsilon_i}\,,\label{V00}
\end{eqnarray}
where $\varepsilon_k=\pm 1$, $k=0,\dots, 3$, and the coefficients
$a_k$ are related to the four charges of the normal form as we
will show. The coset representative of
$\left(\frac{\SL(2,\mathbb{R})}{\SO(2)}\right)^p$ is the product
of $p=4$ copies of (\ref{cosetrep}): $\mathbb{L}_N=\prod_{k=0}^3
e^{\sqrt{2}\,\chi_k\,s_k}\,e^{\phi_k\,H_k}$, where $s_k$ is
represented by the matrix $\left(\begin{matrix}0 & 1\cr 0 &
0\end{matrix}\right)$ on the corresponding $2$-dim. space. The
geodesic with $V(0)=V_0$ is the product of $p=4$ geodesics of the
form (\ref{sol31}),(\ref{sol32}):
\begin{eqnarray}
e^{-\phi_k(\tau)}&=&\mathcal{H}_k\,,\,\,
\chi_0(\tau)=\varepsilon_0\,\frac{\sqrt{2}\,a_0\,\tau}{\mathcal{H}_0}\,,\,\,
\chi_i(\tau)=-\varepsilon_i\,\frac{\sqrt{2}\,a_i\,\tau}{\mathcal{H}_i}\,,\label{solssol}
\end{eqnarray}
where $\mathcal{H}_k\equiv (1-\sqrt{2} \,a_k\,\tau)$ and we have
chosen $\phi_k(0)=\chi_k(0)=0$. Being $\tau<0$ these solutions
will be regular only for $a_k\ge 0$, as we shall assume to be the
case. To uplift these solutions let us first identify the
parameters $a_k$ with $D=4$ charges. To this end we write the
Noether matrix $Q=2\,V_0^T$:
\begin{eqnarray}
{Q}&=&{\sqrt{2}}\,\left(\frac{a_0}{\sqrt{2}}\,
n^{\varepsilon_0}+\sum_{i=1}^3{a_i}\,
n^{-\varepsilon_i}\right)={\sqrt{2}}\,\left({a_0}\,
(H_0+\varepsilon_0\,k_0)+\sum_{i=1}^3{a_i}\,(H_i-\varepsilon_0\,k_i)\right)\,.
\end{eqnarray}
The coefficients $\sqrt{2}\,a_k$ of $H_k$ are scalar (dilatonic)
charges, while the coefficients of $k_k$ are to be identified with
the quantized charges according to eq. (\ref{qr}). In particular,
we can identify $q_0=\varepsilon_0\,a_0$ and
$p^i=\varepsilon_i\,a_i$. As for the fields, the relation of the
three STU dilatonic fields $\varphi_i$ and $U$ to the four
dilatons $\phi_k$ is: $U=\frac{1}{4}\sum_k\phi_k$,
$\varphi_i=\frac{1}{2}\,(-\phi_0-\phi_i+\phi_j+\phi_k)$,
$i,j,k=1,2,3$ \footnote{In the special coordinate frame, where the
prepotential has the form $\mathcal{F}=z_1\,z_2\,z_3$, the three
STU complex scalars read $z_i=b_i-\rmi e^{\varphi_i}$.}. The $D=4$
solution then reads:
\begin{eqnarray}
e^{-2\,U}&=&\prod_k\mathcal{H}_k\,\,,\,\,\,\,e^{2\,\varphi_i}=\frac{\mathcal{H}_0\,\mathcal{H}_i}{\mathcal{H}_j\,\mathcal{H}_k}\,.\label{sol4}
\end{eqnarray}
This implies that
$\mathcal{Z}^0=\chi_0/\sqrt{2}=\frac{q_0\,\tau}{\mathcal{H}_0}$
and
$\mathcal{Z}_i=\chi_i/\sqrt{2}=-\frac{p^i\,\tau}{\mathcal{H}_i}$,
so that equation
$\dot{\mathcal{Z}}=-e^{2\,U}\mathbb{C}\cdot\mathcal{M}_4\cdot\Gamma$
is satisfied. From eq. (\ref{sol4}) we see that, if all the
charges are non-vanishing, there is an attractor mechanism at work
since at the horizon $\tau\rightarrow -\infty$,
$\varphi_i\rightarrow \varphi_i^{(fix)}$, while
$e^{-2\,U}\rightarrow
\frac{A_H}{4\,\pi}\,\tau^2=\sqrt{4\,a_0\,a_1\,a_2\,a_3}\,\tau^2$,
where $A_H$ is the horizon area. The entropy, according to the
Beckenstein-Hawking formula, reads $S=A_H/4=\pi\,\sqrt{|I_4|}$,
where $I_4(p,q)$ is the quartic invariant of the representation
${\bf R}$ of $G_4$ and, on our charges, reads
$I_4(p,q)=4\,q_0\,p^1\,p^2\,p^3=4\,\varepsilon\,a_0\,a_1\,a_2\,a_3
$, where $\varepsilon=\prod_{k=0}^3\varepsilon_k$. The black hole
is regular as long as all the four charges are non vanishing, in
which case one can verify that, by construction, $V_0$ is
nilpotent of order 3 ($V_0^3=0$, $V_0^2\neq 0$) and this is the
maximal degree of nilpotency in $\mathbb{K}_N$. If
$a_0\,a_1\,a_2\,a_3=0$, $V_0$ has a lower degree of nilpotency and
the solution describes a \emph{small} black hole, i.e. a black
hole with vanishing horizon area.  \par We conclude then that
\emph{the generating geodesic of $\mathcal{N}=8$ or of the $STU$
model lifts to the four parameter dilatonic solution}. In what
follows we shall focus on the generating geodesic of regular
solutions, namely for which $\prod_{k=0}^3a_k\neq 0$. The fifth
parameter of the $D=4$ generating solution, which is can be
identified as with the phase $\alpha={\rm
Arg}(Z\,\overline{Z}_1\,\overline{Z}_2\,\overline{Z}_3)$, $Z$ and
$Z_i$ being the STU central and matter charges, can be generated
by a $G$--transformation on the $D=3$ generating geodesic
(\ref{sol4}). Let us now turn to the issue of supersymmetry. The
solution is supersymmetric if $V_0$ is a ${\rm
SL}(2,\mathbb{R})_0$-doublet. Is this a sufficient condition? From
(\ref{V00}) we see that we have $2^4=16$ possible distinct choices
for $V_0$, according to the values of $\varepsilon_k$. We learn
from the analysis in \cite{Bergshoeff:2008be} that the
pseudo-quaternionic structure $\mathfrak{sl}(2,\mathbb{R})_0$
contains the generator $\mathcal{J}=-J_0+\sum_{i=1}^3J_i$.  $V_0$
is an eigen-matrix of $\mathcal{J}$, that is
$[\mathcal{J},V_0]=\pm V_0$, only in the two cases in which
$\varepsilon_k$ are all equal, for which $V_0$ correspond to the
upper or lower component of a ${\rm SL}(2,\mathbb{R})_0$-doublet.
These two choices correspond then to BPS solutions ($1/2$ in
$\mathcal{N}=2$, $1/8$ in $\mathcal{N}=8$). We have $6$ choices
for which $\varepsilon_k$ are not all equal but
$\varepsilon=\prod_{k=0}^3\varepsilon_k=+1$. In these cases, as
shown in \cite{Bergshoeff:2008be}, $V_0$ is no longer eigen-matrix
of $\mathcal{J}$ but it is eigen-matrix of an analogous element on
one of the other three ${\rm SL}(2,\mathbb{R})$ in the isotropy
group. The role of ${\rm SL}(2,\mathbb{R})_0$ is interchanged with
one of the remaining three ${\rm SL}(2,\mathbb{R})$ groups and
this corresponds to interchange the role of the central charge $Z$
with one of the matter charges $Z_i$. The result is a non-BPS
solution in the $\mathcal{N}=2$ model which is still a BPS
solution in the $\mathcal{N}=8$ model (since in the latter the
four ${\rm SL}(2,\mathbb{R})$ in the isotropy group are on an
equal footing, since they form the centralizers of the central
charge matrix $Z_{AB}$ in the normal form). For these non-BPS
solutions $I_4(p,q)>0$. Finally we have $8$ possible choices for
which $\varepsilon=\prod_{k=0}^3\varepsilon_k=-1$. In these cases
$V_0$ does not factorize at all and $I_4(p,q)<0$. The
corresponding solution is non-BPS. On the generating geodesic then
the only choices which lift to a BPS solution are those in which
$V_0$ is a ${\rm SL}(2,\mathbb{R})_0$-doublet. \par We have found
then three classes of choices for $V_0$ yielding different kinds
of solutions in $D=4$. These classes are mapped into one another
by transformations of the form $S^{-1}\,V_0\,S$ whose effect is to
switch the grading, $n^{\varepsilon_k}\rightarrow
n^{-\varepsilon_k}$, for a subset of the $p=4$ terms in eq.
(\ref{V00}), keeping the coefficient of $n^{\varepsilon_k}$
positive. The matrix $S_k$ which does the job on each term is
$S_k=e^{\frac{\pi}{2}\,k_k}\,e^{{\rmi \,\pi}\,J_k}$ since
$S_k^{-1}\,n^{\varepsilon_k}\,S_k=n^{-\varepsilon_k}$. The matrix
$S$ which switches the grading to a number $\ell$ of terms is
then:
\begin{eqnarray}
S&=&\prod_{i=1}^\ell S_{k_i}\,.\label{wick}
\end{eqnarray}
If $\ell$ is odd, $S$ will cause
$\varepsilon=\prod_{k=0}^3\varepsilon_k$ to change sign and thus
will map a $\mathcal{N}=8$ BPS solution $I_4(p,q)>0$ into a
non-BPS one $I_4(p,q)<0$. Since $S\in
G^{\mathbb{C}}/G=E_{8}^\mathbb{C}/E_{8(8)}$, the orbits of the two
kind of geodesics will be different real sections of a same
$E_{8}^\mathbb{C}$-orbit, consistently with \cite{Bossard:2009we}
. This conclusion is a direct consequence of the analysis in
\cite{Bergshoeff:2008be}, which we have reviewed in the present
section, about the relation between the generating geodesics of
BPS and non-BPS black holes.

\section{Conclusions} \label{sec:concl}
In this note we have pursued a different avenue  to study the
exact integration of all the spherically symmetric black hole
solutions to supergravity theories with symmetric target spaces.
This approach applies to all supergravities with more than 8
supercharges and to an interesting subset of  theories with 8 and
less supercharges. Our treatment is not referring to (hidden)
supersymmetries of the theory, and uses the equivalence between
the equations governing the radial evolution of the fields in four
dimensions,
and the geodesic motion of a particle on an appropriate pseudo-Riemannian symmetric space. \\
\indent We established that  the purely $t$-dependent backgrounds,
which reduce to those of a one-dimensional sigma model, admit a
Lax pair representation and are fully integrable. The integration
algorithm we have exploited depends more fundamentally on the
diagonalizability of the initial data. We leave the details of the
analysis for extending the algorithm
to non-diagonalizable initial conditions for an upcoming paper \cite{Chemissany}. \\
\indent We were able to show explicit analytic formulae for the
general integral of simple examples like the
$\SL(2,\mathbb{R})/\SO(1,1)$ and the  $\SL(3,\mathbb{R})/\SO(2,1)$
models.  The main message of this note is that the integration
algorithm is fully explicit and hopefully we have been clear
enough to illustrate this fact. We have applied this analysis to
construct the \emph{generating geodesics} corresponding to regular
and small $D=4$ black holes, which is shown to  belong to a
submanifold consisting of a direct product of $dS_2$ spaces. We
have focused on
 maximal supergravity and its STU truncation and, as a byproduct, we
have written the explicit form (\ref{wick}) of the ``Wick
rotation'' mapping the initial data $V_0$ of BPS and non-BPS
regular solutions in these models.\par As a final comment we
emphasize that  formulating the problem in terms of a Lax pair
equation is important for ultimately proving the complete
Liouville integrability (i.e. the global existence of a number of
constants of motion in involution equal to the number of scalar
fields) of the $D=3$ model, at least for symmetric spaces. This
however is still an open issue which we are also working on in
\cite{Chemissany}.

\section{Acknowledgements}
We would like to thank P. Fre', Y. Kodama and A. Sorin for
enlightning discussions. W.C. is supported in part by the Natural
Sciences and Engineering Research Council (NSERC) of Canada. The
work of J.R. and M.T.  is supported
in part by the Italian MIUR-PRIN contract 20075ATT78. T.V.R. is supported by the G\"oran Gustafsson Foundation and
he likes to thank the Politecnico di Torino for its hospitality.



\end{document}